          [2001/04/25 1.1 (PWD)] 
\documentclass[a4paper,twocolumn]{esapub} 

\usepackage{times} 
\usepackage{natbib} 
\usepackage{graphicx} 
 
\title{THEORETICAL ASPECTS OF ASTEROSEISMOLOGY: SMALL STEPS TOWARDS A GOLDEN FUTURE} 
\author{Maria Pia Di Mauro} 
\affil{INAF- Osservatorio Astrofisico di Catania, Via S. Sofia 78, 95123
Catania, Italy}

\begin{document} 
 
 
\maketitle 
 
\begin{abstract} 

The  current status of asteroseismic studies is here reviewed and the 
adequate techniques of analysis available today for the study 
of the oscillation frequencies are presented. 
Comments on prospects 
for future investigations through the possibility of getting
ever more precise asteroseismic observations from ground and space
are given.
\end{abstract} 

\section{Introduction} 
The present conference took place at a period 
which marks a significant milestone in the development of stellar physics:
asteroseismology, which aims to infer the structural properties of stars
which display multi-mode pulsations, 
 has entered in a new golden age.
In fact, ground-based observations have reached such level of accuracy to allow investigation of several type of oscillating stars; the satellite MOST (Walker et al. 2003), the first space mission totally dedicated to the observation of stellar oscillations, has been successfully launched in June 2003 and has just
started to reveal primary results on the analysis of pulsations seen outside earth's atmosphere.
In order to make full use of these observational successes,
several theoretical techniques have been developed or adopted from helioseismology to probe the internal characteristics of stars other than the Sun. 

Here I will review on the theoretical aspects of asteroseismology, considering
 methods and tools available today to manipulate observed frequencies of oscillation in order to investigate the evolutionary and structural properties of the stars.
Example of application to real data set will be given for each asteroseismic
diagnostic tool considered.
Some recent results will be shown and comments will be given on perspectives
for the future.
For basic concepts on the theory of stellar oscillations 
I refer to classical books (e.g. Cox 1980; Unno et al. 1989); for theoretical 
 methods and developments in asteroseismology see reviews by, e.g.,  
 Christensen-Dalsgaard (2003, 2004), 
Christensen-Dalsgaard and Dziembowski (2000), Dziembowski (2003b), Gough (2003),
Patern\`o, Di Mauro and Ventura (2003), Roxburgh (2002, 2004).

\section{From Helio- to Asteroseismology}

During the last decades, 
numerous observational and theoretical efforts in the study of the acoustic modes of solar oscillations,
has brought to a detailed knowledge of
the interior of the Sun.

The experience acquired in helioseismology on handling pulsations frequency data provides a good starting point for asteroseismic investigations.
However, inferences
of the interior of stars other than the Sun 
appear to be much more complicated and less outstanding in terms of 
achievable results.
The large stellar distances, the point-source character of the stars, the
 low amplitude of the oscillations and the effect of the earth's atmosphere on the signal, restrict the asteroseismic studies
to the use of small sets of data often characterized by modes with only low harmonic 
degrees ($l\leq 4$).
However, the main difference between helio- and asteroseismology is that global parameters of the Sun, are much better known than they are for any other star.
Luminosity, effective temperature, surface composition and $v \sin i$ are 
obtained, within a certain error, from spectrum analysis of distant stars;
age and composition are estimated approximately only for stars in clusters;
masses and radii are measured only for spectroscopic binaries.
In addition, all these measurements 
can be affected by unknown effects such as loss, 
accretion, or diffusion of mass.
Thus, the structure of the model  of the 
stars cannot be so well constrained such as that of the Sun.

All these problems make asteroseismic inferences
quite difficult, so that
the successes reached by helioseismic studies are beyond the possibility of asteroseismology.
Nevertheless, several attempts have been made, during the last years, 
with the aim to identify oscillations in distant stars
and also to model the 
stellar pulsational phenomena.
 In fact, oscillations have several advantages over 
all the other observables:
pulsational instability has been detected in 
stars in all the
evolutionary stages and of different spectral type;
frequencies of oscillations can be measured with high accuracy and
 depend in very simply way on the equilibrium structure of the model (only on two independent variables, e.g. the local adiabatic speed of sound and density); different modes are confined and probe different layers of the interior of the stars.
Thus, accurate observations of the acoustic frequencies, 
can be used not only to study the physics of the stellar interior,
but also
to constrain theories of stellar evolution.
 
\section{Pulsating stars}

\begin{figure} 
\centering 
\includegraphics[width=0.9\linewidth]{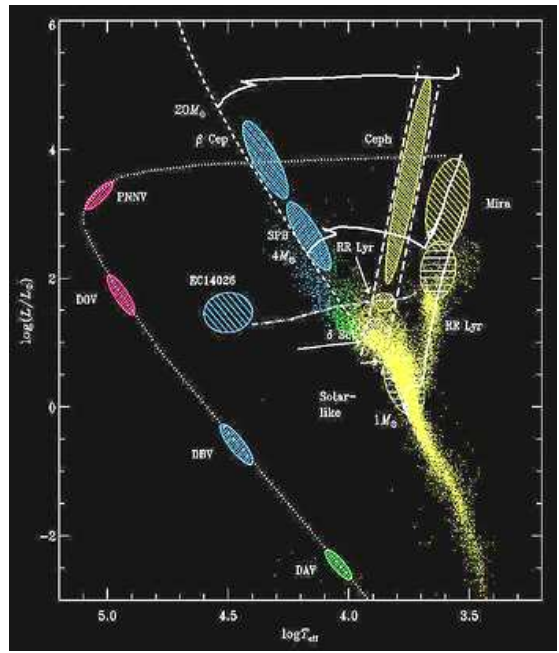} 
\caption{HR diagram showing several classes of pulsating stars. Courtesy of J. Christensen-Dalsgaard.} 
\label{HRall} 
\end{figure}

The historical pulsating stars, like the classical Cepheids, W-Virginis and RR-Lyrae, characterized by a large luminosity variations, show only one or two pulsation modes, limiting the knowledge of the structural properties to the mean density or in better cases to the mass and radius. 
Asteroseismology is understood as the study of pulsations on stars in 
which many oscillation modes are excited at once.
This definition is broad enough to include many types of pulsating stars.
Multi-mode oscillations have been detected in stars 
 which are spread 
over a significant part of the HR-diagram, reflecting different evolutionary stages from main-sequence to the white dwarf cooling sequence.
 Luminosity and effective temperature of the major classes of pulsating stars are shown
 in the HR-diagram of Fig.~\ref{HRall}.

We can divide the stars on which it is possible to apply seismic techniques
into two groups. A group of stars show oscillations with intensity amplitude in the range of millimagnitudes and includes white dwarf,  $\delta$ Scuti stars, the rapidly rotating Ap stars, the $\beta$-Cephei stars, the slowly pulsating B (SPB) stars  and the
$\gamma$ Doradus stars. Stars of these classes
 show modes excited 
by $\kappa$-mechanism due to a perturbation of the opacity, caused by structural conditions typical of each considered class.
The second group includes 
the stars with tiny oscillations amplitude (around $10 \,\mu$magnitudes or less) and includes the so-called solar-like stars, in which, like the Sun,
oscillations are excited by turbulent convection 
in the outer convective envelope.

For the purposes of this review, however, I will concentrate manly on solar-type stars and I will give just a summary on some theoretical results
obtained by asteroseismology on some other pulsating targets at the end of this review.
The principal reason for doing so is due to the excitement of the
 community for the recent observational successes obtained on some
 solar-type stars (e.g., $\alpha$-Can Min and $\alpha$-Cen) which have given
the possibility, for the first time, of handling 
large sets of accurate oscillation
 frequencies to investigate the internal structure of the stars.
Moreover, it should be pointed out, 
 that
 methods and techniques developed and adapted for 
solar-type stars,
 are also valid for the other pulsating stars.
A detailed and updated description of the characteristics of the main
asteroseismic targets can be found in recent
reviews (e.g. Dziembowski 2003b; Kurtz 2004; Patern\`o, Di Mauro and Ventura 2003).

\subsection{Properties of pulsations} 

 Many of the oscillation modes which can be detected in stars are limited
to only the lowest degree ones, owing to the point-like character
of these sources.
If the condition that $l\ll n$ is satisfied, the excited p-modes
can be described in terms of the asymptotic theory (Tassoul 1980), 
which
 predicts that oscillation frequencies $\nu_{n,l}$ of acoustic modes,   
 characterized by radial order $n$, at harmonic degree $l$   
should satisfy the following approximation:   
\begin{equation}   
\nu_{n,l}=\Delta\nu\left(n+\frac{l}{2}+\alpha+\frac{1}{4} \right)   
+\epsilon_{n,l} \; ,   
\label{eq1}   
\end{equation}   
where $\alpha$ is a function of the frequency determined by the
properties of the
surface layers, $\epsilon_{n,l}$   
 is a small correction which depends on the conditions in the stellar core.
$\Delta\nu$ is the inverse of    
the sound travel time across the stellar diameter: 
\begin{equation}
\Delta\nu={\left(2\int_{0}^{R}\!\frac{{\rm d}r}{c}\right)}^{-1}, 
\end{equation}
where $c$ is the local speed of sound at radius $r$ and $R$ is the photospheric stellar radius.
The property expressed by Eq.~(\ref{eq1}) may provide almost immediate asteroseismic inferences on stellar parameters and constraints on theoretical models for a variety of solar-like stars in a wide range of evolutionary stages.   
For a given $l$, the acoustic spectra
show a series of equally spaced peaks between p modes of same degree and adjacent $n$, whose frequency separation represents 
the so called large separation which is
approximately equivalent to
$\Delta\nu$:
\begin{equation}
\Delta\nu \simeq\nu_{n+1,l}-\nu_{n,l}\equiv\Delta\nu_{l}\,.
\label{EQ_3}
\end{equation}
 The spectra are characterized by another series of peaks, whose narrow separation is $\delta\nu_{l}$, known as small separation:
\begin{equation}
\delta\nu_{l}\equiv \nu_{n,l}-\nu_{n-1,l+2}=(4l+6){\rm D}_{0}
\label{EQ_4}
\end{equation}
where
\begin{equation}
{\rm D}_{0}=-\frac{\Delta\nu}{4\pi^{2}\nu_{n,l}}\left[\frac{c(R)}{R}-
\int_{0}^{R}
\frac{{\rm d}c}{{\rm d}r}\frac{{\rm d}r}{r}\right] \, .
\end{equation}
Instead of the small separations one could consider the so-called 'second differences' of frequencies:
\begin{equation}
\delta_2\nu_{l}\equiv \nu_{n+1,l}-2\nu_{n,l}+\nu_{n-1,l}\,.
\label{EQ_10}
\end{equation}

$\Delta \nu$, and hence the general spectrum of acoustic modes,   
scales approximately as the square root of the mean density, that is,   
for fixed mass, as $R^{-3/2}$.   
 On the other hand, the small frequency separation and the second difference,   are both sensitive to the chemical composition gradient in   
 central regions of the star and hence to    
 its evolutionary state. 
Thus, the determination of both   
large and small frequency separation, $\Delta\nu_l$ and $\delta \nu_l$,   
provides measures of the mass and the age of the star 
(e.g.\ Christensen-Dalsgaard 1988).
See review by Gough (2003) on the potential of
frequencies separations as asteroseismic diagnostic.   

Analogous properties have been derived for the g-modes.
Tassoul's theory shows that in the asymptotic regime the g-modes are nearly uniformly spaced in period.
g-modes will be discussed in more details in Sec. \ref{S.WD}, since the only types of stars in which these are commonly observed are rather special cases.

  \subsection{Sub-surface effects and a new seismic indicator}

Seismic analysis of observed acoustic frequencies is an extremely powerful tool for the investigation of the internal structure of the stars, but the use of large and small separations can be misleading if not accurately considered.
Theoretical pulsation frequencies, essential for asteroseismic investigation, are calculated on theoretical models of stars which are inevitably affected by errors. In particular, the structure of the near-surface regions of stars is quite uncertain: there are still substantial ambiguity in modelling the convective flux, considering the sources and the mechanisms of excitation and damping of the oscillations, defining an appropriate equation of state to describe the thermodynamic properties of the stellar structure, as well as in the treatment of non-adiabatic effects. Limiting the investigation to the use of small separations do not solve the problem. In fact,
 the small separation which  is determined principally by condition in the 
core, also retains some sensitivity to the mean density and to the detailed properties of the stellar envelope.

To solve these problems the use of a new seismic indicator was introduced by
 Roxburgh and Vorontsov (2003a), which compared pulsating properties 
of several models with
exactly the same interior structure, but different outer envelopes,
obtained by keeping constant the first adiabatic exponent all along the radius
 or by introducing linear variation of the polytropic index in the structure of the envelope. They showed that the ratios of small to large separations,
\begin{equation}
r_{l}=\delta\nu_{l}/\Delta\nu_{l}\, ,
\end{equation}
are independent of the structure of the outer layers and hence
can be used as diagnostic of the interior of stars.

\section{Solar-type stars}

\begin{figure} 
\centering 
\includegraphics[width=0.9\linewidth]{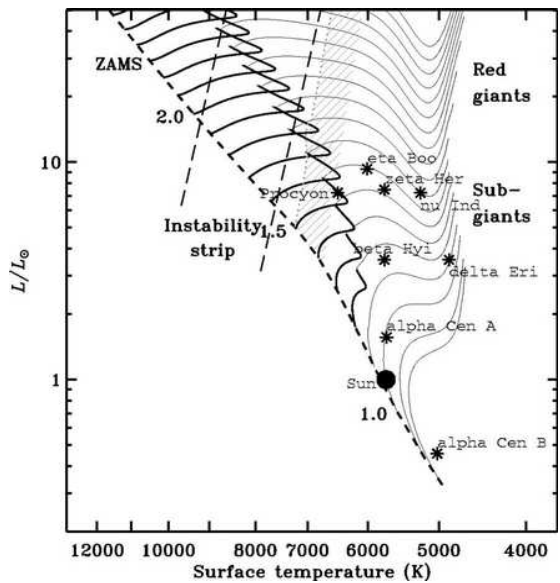} 
\caption{HR diagram showing location in luminosity and effective temperature of the solar-type pulsating stars. Asterisks indicate targets on which solar-type oscillations have been already detected. The Sun is shown as a full circle. The solid lines are evolutionary tracks calculated for increasing values of mass. Courtesy of G. Houdek.} 
\label{HRsolartype} 
\end{figure}

Solar-type stars are F, G and possibly K main-sequence and sub-giants stars in which oscillations are excited stochastically by vigorous near-surface convection in a broad spectrum of low amplitude p-modes, as in the Sun. There is also good evidence, both from photometry with the WIRE satellite (Buzasi et al. 2000) and ground-based velocity studies, for solar-like oscillations in G, K and semi-regular M giants. The oscillation periods are expected to be in the range from the typical 5-min (main-sequence stars), as in the Sun, up to about a few days (sub-giant and giant stars). Fig.~\ref{HRsolartype} shows location in
HR-diagram of the solar-type pulsating stars and of some selected targets.

\subsection{Amplitudes and life-time of solar-like oscillations}

One of the greatest deficiencies in modelling oscillations in stars with 
surface convection zones is the lack of a proper theory to describe 
convection and its interaction with pulsations. Several attempts have
 been made in recent years to address this problem.
 Christensen-Dalsgaard and Frandsen (1983) made the first effort to give 
a rough prediction of amplitudes of solar-like oscillations. 
They found velocity and luminosity amplitudes increasing with age and 
with mass along the main sequence. 
Simple scaling laws to 
estimate the amplitudes in velocity have been proposed by, e.g., 
Kjeldsen and Bedding (1995, 2001).

Impressive progress has been made on hydrodynamical simulations of convection, and its interaction with pulsations, leading
Houdek et al. (1999) to give estimates
 of the pulsation properties, including the damping rates, of main-sequence stars by using models based on  
mixing-length treatment of convection.
According to theoretical estimations (e.g. Houdek et al. 1999, 
Kjeldsen and Bedding 1995) the expected luminosity variations are of about a
 few mmag in giant stars while in main-sequence and sub-giant stars are 
so small (from $10^{-3}$ up to a few $10^{-2}\,{\rm mmag}$) to be 
well below the detection limit for ground-based photometric observations.
The radial velocity amplitudes are expected to be at most 
$50-60\,{\rm m\,s^{-1}}$ in K giants, $1-2\,{\rm m\,s^{-1}}$ in 
F and G sub-giants and even smaller in main-sequence solar-type stars. 
These amplitudes are at the edge of the 
photon noise limits of most of the available spectrographs.
 The theoretical prediction have been tested and compared with observational 
results, revealing a fairly good agreement for the several solar-like stars
already detected.
However, calculation of damping rates are still a problem. Theoretical prediction  have
been able to reproduce
 observed lifetimes of oscillations  
only in the case of the Sun.
Observations of stars which are somewhat more evolved than the Sun, 
like Procyon (Barban et al. 1999), $\alpha$~Cen A (Bedding et al. 2004),
$\beta$ Hyi (Bedding et al. 2002)
and $\xi$~Hya (Stello et al. 2004)
have
 shown mode lifetimes shorter than the values 
predicted by the theoretical models (Houdek et al. 1999). 
This inconsistency with observations seems to indicate that 
there is still some contributions to damping so far ignored
 in the theory.

\section{Stellar Modelling} 

 Reliable calculation of stellar models
is the basic prerequisite
for all asteroseismic investigations of stellar internal
properties.
The structure models are produced 
 by employing up-to-date physical information and the most updated
 input parameters (composition, mass, mixing-length) 
in order to reproduce all the spectroscopic and photometric observations of
the considered stars.
Evolutionary    
sequences are calculated  
by including additional effects such as overshoot from the convective core 
(e.g. Di Mauro et al. 2003) during the main-sequence phase, diffusion and settling
 of helium and heavy elements (e.g. Vauclaire 2003).

In the case of solar-type stars, detailed classical models have been produced on
several well observed targets:
$\alpha$ Cen (e.g. Guenther and Demarque 2000; Morel et al. 2000, Th\`evenin et al. 2002; Thoul et al. 2003; Eggenberger et al. 2004b); $\eta$ Boo (Christensen-Dalsgaard, Bedding and Kjeldsen 1995; Guenther and Demarque 1996; Di Mauro et al. 2003, 2004; Kervella et al. 2003; Guenther 2004), 
Procyon A (e.g. Barban et al. 1999; Chaboyer, Demarque and Guenther 1999; Di Mauro, Christensen-Dalsgaard and Weiss 2000, Kervella et al. 2004, Provost, 
Marti\'c and Berthomieu these proceedings), $\beta$ Hydri (Di Mauro, Christensen-Dalsgaard and Patern\`o 2003; Fernandes and Monteiro 2003;
Dravins, Lindegren and Vandenberg 1998); $\xi$ Hydrae (Teixeira et al. 2003; Christensen-Dalsgaard 2004).

Unfortunately, modelling of distant stars involves physical effects which substantially complicates the calculation.    
One of the most crucial problem in computing stellar structure 
 is related to the modelling of
the convective transport of energy in the interior and the associated 
physical effects such as the evolution of the composition in presence of convective regions (e.g. Dupret, these proceedings ).
See the reviews by Demarque and Robinson (2003) and D'Antona, Montalban and Mazzitelli (2004) about sophisticated models of convection.
 
Another trouble is represented by the fact that effects of rotation in 
stellar modelling are often neglected.
The rotation  may affect oscillation frequencies forming
 rather complex power spectra.
In order to identify and reproduce the observed modes it is necessary to develop an appropriate treatment of rotation in the evolutionary models and in the calculation of oscillation frequencies.
 To date, rotational effects have been taken into account mainly 
using perturbation theory approaches. The Coriolis and the centrifugal forces, which in the case of moderate and fast-rotators cannot be neglected, have been taken into proper account as first-, second- and even third order
 perturbations. Many authors (Saio 1981; Gough and Thompson 1990; Dziembowski and Goode 1992; Soufi, Goupil and Dziembowski 1998; Karami,
Christensen-Dalsgaard and Pijpers 2003) have approached the problem under various approximations, also including radial and latitudinal dependence of rotation and
magnetic fields,  which remove the degeneracy of  the modes.
However
perturbation theory might be inadequate for the most rapidly rotating stars.

Calculation of the evolution and pulsations of two- or three-dimensional models, represents a solution for all these problems and turns out to be strongly recommended,
but at present, very difficult to be attained (e.g. Demarque and 
Robinson 2003).
\begin{figure*}
\centering
\includegraphics[draft=false,scale=0.27,angle=270]{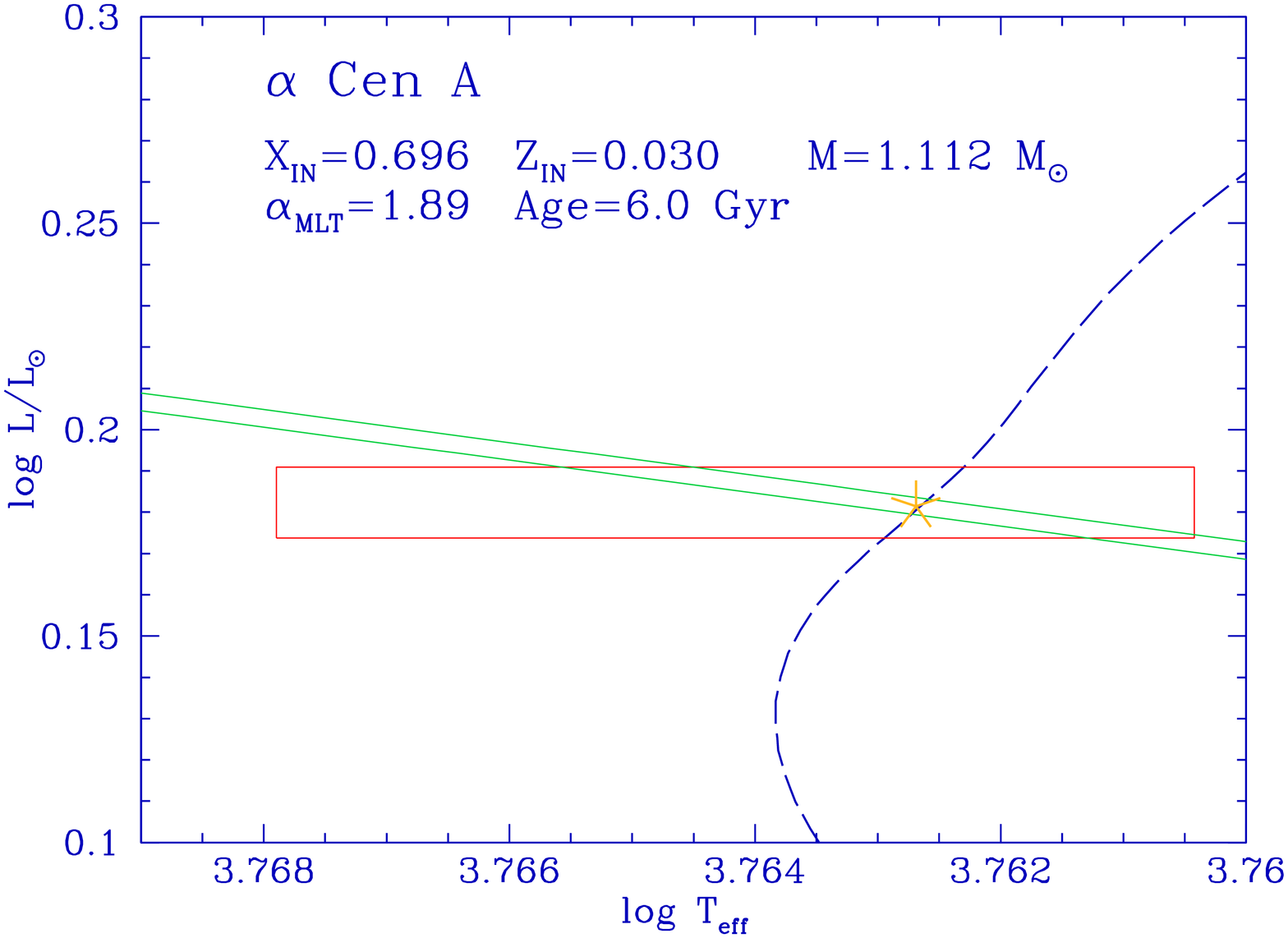}
\includegraphics[draft=false,scale=0.27,angle=-90]{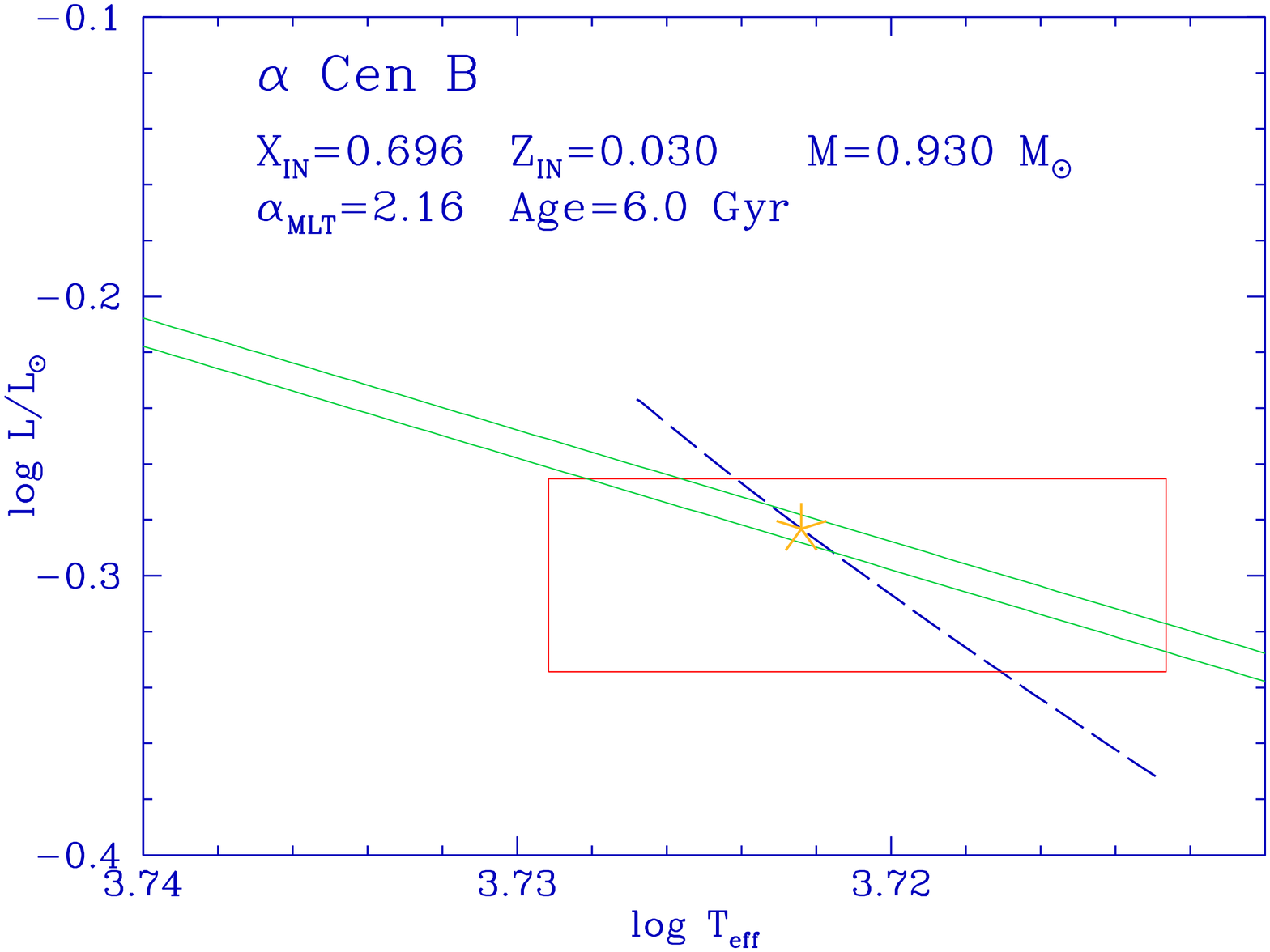}
\caption{The dashed blue lines show evolution tracks plotted in HR-diagrams 
for $\alpha$ Cen A and B, respectively on the left and right panel.
 The tracks have been calculated with the Li\'ege evolution code by A. Miglio and J. Montalban by assuming OPAL EOS.
 Input values of masses, age and mixing length parameters ($\alpha_{\rm MLT}$) are indicated.
The rectangles define the one-sigma error box for   
the observed luminosities and effective temperatures and the green lines show 
the one-sigma error box for   
the observed radius.
The symbol star indicates the location of the selected models. 
Courtesy of A. Miglio.}
\protect\label{F.1}
\end{figure*}
\begin{figure}[h]
\centering
\includegraphics[draft=false,scale=0.44]{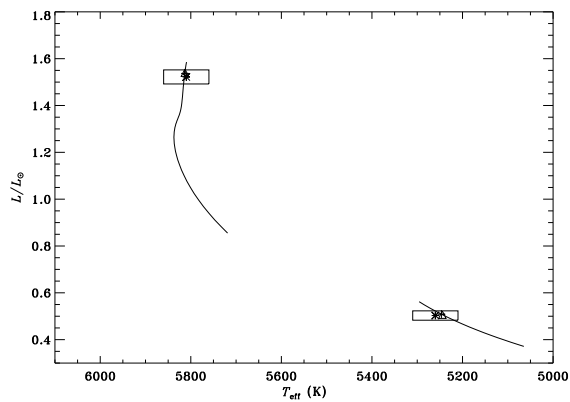}
\caption{Evolution tracks for $\alpha$ Cen A and B plotted in a HR-diagram 
calculated with the evolution code of Christensen-Dalsgaard, with $M_A=1.1095\,M_{\odot}$ and $M_B=0.9301\,M_{\odot}$, the OPAL EOS and an age of $7.382$ Gyr. The rectangles define the one-sigma error boxes around    
the observed luminosities and effective temperatures (asterisks).
 The triangles indicate the location of the selected models. Courtesy of J. Christensen-Dalsgaard.}
\protect\label{F.2}
\end{figure}
\subsection{Model fitting method and application to the $\alpha$ Cen system} 

One of the most recent strategy introduced in order to find the models
which better reproduce the observations is represented by 
an optimization procedure known as 'model fitting' method.
Given a set of input parameters, such as mass, mixing-length parameter and composition of a star, the method consists in searching, among all the possible models, the one
 which best fits all the known observables. This can be obtained
by performing a $\chi^2$ minimization, looking for the parameters which minimizes:
\begin{equation}
\chi^2=\sum_i^N\left(\frac{O_i^{mod}-O_i^{obs}}{\sigma_{i}}\right)
\end{equation}
where $O_i^{obs}$ are all the observables - luminosity, effective temperature, radius, oscillation frequencies, large and small separations, ratio of small to large separation - and $O_i^{mod}$
are the theoretical values calculated on the models. 
 The observed errors of the parameters are $\sigma_i$.
Extensive explanation of the method can be found in Guenther and Brown (2004)
and Roxburgh (2004).
Although in principle the method might represent an useful strategy,
it still presents some problems: the solution depends on the initial inputs and therefore it does not explore the global space of parameters.
To solve this problem , Metcalfe et al. (these proceedings) have suggested
to use genetic algorithms,
well developed and applied for modelling the white dwarfs, also for the case of the solar-type stars .
\begin{table}
 \begin{center} 
\caption{Mass, luminosity, effective temperature, radius and
logarithmic value of the iron abundance
 relative to the solar one
 for the two solar-type components of the $\alpha$ Cen system, adopted from Eggenberger et al. (2004b).} 
\vspace{1em} 
\renewcommand{\arraystretch}{1.2} 
\begin{tabular}{lcc}\hline 
  &
$\alpha$ Cen A&
$\alpha$ Cen B\\  \hline
$M/M_{\odot}$ & 
   $ 1.105\pm0.007$&
    $0.934\pm0.006$\\  
$L/L_{\odot}$&    
$1.522\pm0.030$&
$0.503\pm0.020$\\ 
   
$T_{\rm eff}(\mathrm{K})$&    
$5810\pm50$&
$5260\pm50$\\   
$R/R_{\odot}$&
$1.224\pm0.003$&   
$0.863\pm0.005$\\   
   
[Fe/H]&  
$ 0.22\pm0.05 $&    
  $0.24\pm 0.05 $\\  
\hline    
\end{tabular}  
\label{T.1}
  \end{center} 
\end{table}   

'Model fitting' is particularly valuable when applied to a set of stars in a cluster or in a binary system: the increase of
 parameters to be fitted,
 by assuming same age and composition for all the components of the system,
 enhance the quality of the minimization technique.
An immediate application of this optimized procedure has been performed, for example, 
in the case of the $\alpha$ Cen system by several authors (e.g. Eggenberger et al. 2004b; Guenther and Brown 2004).
Here, I present the 
results not yet published obtained by A. Miglio and J. Montalban 
which implemented the Code Li\'egeois d'Evolution Stellaire, 
and by J. Christensen-Dalsgaard,
 whose code was implemented by T. Teixeira.
Many spectroscopic measurements of both components of the $\alpha$ Cen 
system can be found in literature; for a summary see Morel et al.~(2000)
and Eggenberger et al.~(2004b).
Taking into account the numerous observational constraints 
listed in Table~\ref{T.1}, including new seismological observations for $\alpha$ Cen A (Bedding et al. 2004) and $\alpha$ Cen B (Carrier and Bourban 2003)
a common solution was obtained by the
two groups, and respectively shown in Figs.~\ref{F.1} and \ref{F.2}.
The results
 indicate that a best fit with all the available observations
can be obtained by assuming a mass of about $M_A=1.11M_{\odot}$ for $\alpha$ Cen A  and of about $M_B=0.93M_{\odot}$ for
 $\alpha$ Cen B, identifying both components as being in the
 main-sequence phase. The composition was assumed such that the initial hydrogen abundance was $X=0.7$ and the metallicity $Z=0.03$.

\section{The potential of the mixed modes}
   
The properties of the solar-like oscillations are expected to change
as the stellar structure evolves as a consequence of the
hydrogen exhaustion in the core.  
According to Eq.~(\ref{eq1}),   
oscillation frequencies of a given harmonic degree should decrease as the star evolves and 
the radius increases   
and should be almost uniformly spaced by $\Delta \nu$ at   
each stage of evolution.    
However, while the   
radial modes seem to follow closely Eq.~(\ref{eq1}),   
the frequencies of some non-radial modes appear to increase    
suddenly at certain stages of evolution   
(Christensen-Dalsgaard, Bedding and Kjeldsen 1995; Guenther and Demarque 1996). 
This  is due of the occurrence to the so called 'avoided crossing'.
The core contraction,
as the star evolves and the radius expands,
causes an increase of the local gravitational acceleration
and of the gradients in hydrogen abundance, and hence of
the buoyancy frequency in the deep interior of the star. 
As a consequence g modes with high frequencies
are allowed to propagate and can interact with a p mode of similar frequency 
and same harmonic degree, giving rise to a mode with mixed character, which behaves as a g mode in the interior and as a p mode in the upper layers.
The interaction can be explained as the coupling of two oscillators of similar frequencies.
The effect of 
the coupling becomes much weaker for modes with higher harmonic degree,
since in these cases the gravity waves are better trapped in the 
stellar interior and better separated from the region of propagation 
of acoustic waves.

\subsection{Mixed modes in $\eta$ Bootis: main-sequence or subgiant star?}
\begin{table*}
\caption{Relevant parameters,
the mass $ M$, the $Age$, the luminosity $L$, the effective temperature $T_{\rm eff}$, 
the surface radius $R$, the overshooting parameter $\alpha_{\rm ov}$ and
the location  
$r_{\rm cb}$ of the base of the convective envelope    
in units of $R$, for two   
models of $\eta$ Boo, computed with the OPAL EOS 
and $Z=0.04$. Model~1 and 2 represent evolutionary structure respectively  
in post-main and main-sequence phases.}\label{T.2} 
\vspace{1em} 
\renewcommand{\arraystretch}{1.2} 
\begin{center}
\begin{tabular}{cccccccc}\hline 
 Model &  
$M/M_{\odot}$ & 
$Age$ (Gy) &
 $L/L_{\odot}$ &    
 $T_{\rm eff} \; (\mathrm{K})$ &    
 $  R/R_{\odot}$ &    
 $  \alpha_{\rm ov}$ &
$r_{\rm cb}/R$ \\ \hline   
  1 &
$1.71$ &   
 $2.44$ &    
  $9.07$ &    
  $6072$ &    
 $2.73$  &   
 $0.1$ &
$0.85$ \\   
   
  2 &
 $1.82$ &   
 $1.96$ &    
  $8.93$ &    
  $6009$ &    
 $2.76$  &   
 $0.2$ &
$0.84$ \\   
   
\hline    
\end{tabular} 
\end{center}  
\end{table*}   
\begin{figure*}[htb]   
\centering  
\vspace{-0.8cm}
\includegraphics[draft=false,scale=0.40]{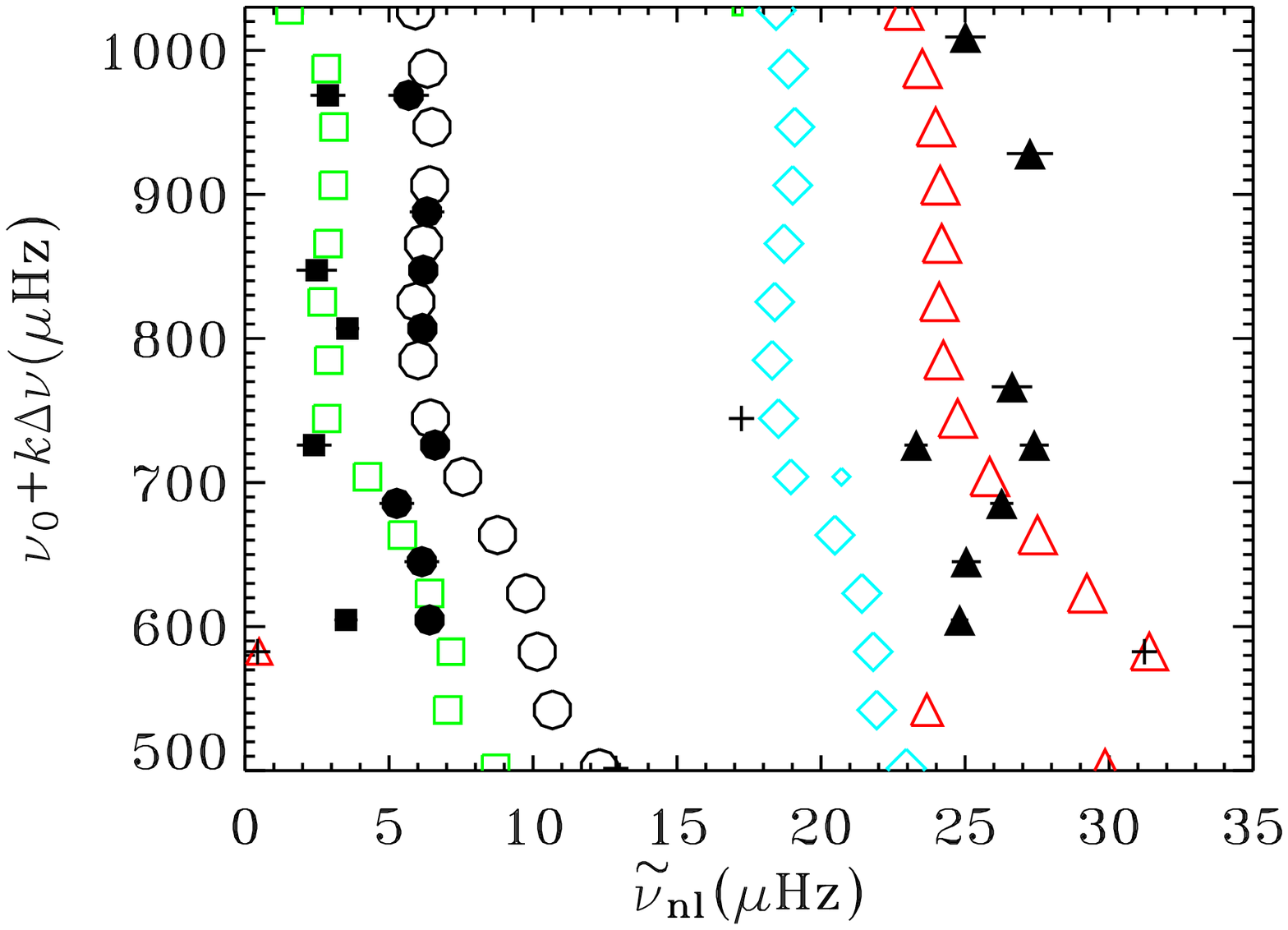}
\includegraphics[draft=false,scale=0.40]{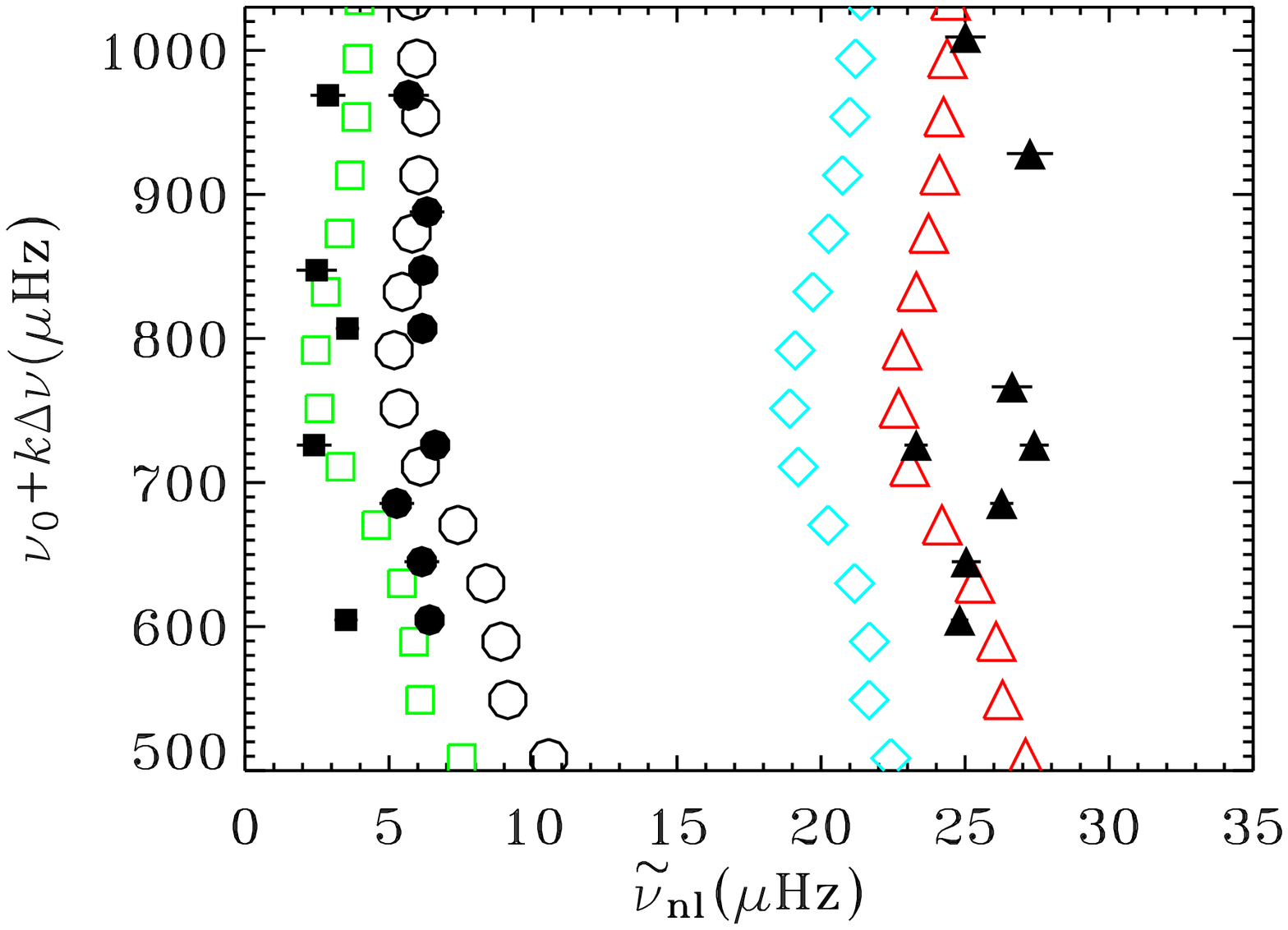}
\caption{Echelle diagrams for 
Model~1 (left panel) and Model~2 (right panel) of Table~\ref{T.2}, 
which are respectively in the post-main-sequence and
 main-sequence phases. 
The open symbols represent the computed frequencies.
The filled symbols with error bars show observed frequencies and 
errors (Kjeldsen et al., 2003).
Circles are used for modes with $l=0$, triangles for $l=1$, squares
for $l=2$, and diamonds for $l=3$.   
Theoretical and observed frequencies are 
plotted with $\Delta \nu=40.47 \, \mu\mathrm{Hz}$.   
 The size of the open symbols indicates the relative
surface amplitude of oscillation of the modes.
Crosses are employed for modes with small predicted   
amplitude (e.g. g modes). For details see Fig.~4 of Di Mauro et al. (2004).}
\protect\label{F.4}   
\end{figure*}   
As an example of the use of mixed modes as asteroseismic diagnostic in the 
stars, I consider the case of $\eta$ Bootis.
In Di Mauro et al. (2004) authors built a grid of stellar evolutionary sequences tuned to match the position of 
$\eta$ Boo in the theoretical HR diagram, considering
the inclusion of convective overshooting, and testing the 
use of different equations of state
and of different treatments of convective transport.
They concluded that present observations
 are consistent with 
two possible evolutionary scenarios for   
$\eta$ Boo:   
(i) a subgiant star in post-main-sequence phase, whose oscillation spectrum contains frequencies
of nonradial modes with mixed character due to avoided crossings;   
(ii) a less evolved star in the main-sequence phase
which show p modes with no mixed character and frequencies which
 follow the asymptotic theory.  
Fig.~\ref{F.4} shows
the echelle diagrams of the observed frequencies in comparison with
 the theoretical frequencies computed for
Model~1 and 2 of Table~\ref{T.2}, respectively 
in post-main-sequence and 
main-sequence phases.   
At this stage it can be concluded that
the accuracy of the existing frequency observations does not allow a clear   
distinction between the two proposed scenarios, although
it is evident, but hardly decisive, 
that the post-main-sequence phase 
is a rapid phase of the evolution, which might be less likely to be observed.
For the future there is hope that more accurate    
observations will help to detect presence or absence of mixed modes and then to identify uniquely 
the present evolutionary state 
and clarify the   
properties of the interior of this star.  

\subsection{Mixed modes in the red giant $\xi$ Hya}

The analysis of the properties of solar-type oscillations has so far confined to stars which lie in the main sequence or subgiant phase.
However there have been reports of detections of solar-like oscillations
also in giant stars, and in particular in few red giants (e.g. 
Buzasi et al. 2000; Merline 1999, Retter et al. 2003; Frandsen et al. 2002; Stello et al. 2004; Retter et al. 2004; Barban et al. these proceedings).
Although there are difficulties in the identification of 
the observed modes 
and severe problems of theoretical interpretation, 
Guenther et al. (2000) and Dziembowski et al. (2001) for $\alpha$ Uma
and Teixeira et al. (2003) and Christensen-Dalsgaard (2004) for $\xi$ Hya
succeed to draw important
 conclusions on the structural and pulsational properties of red giant stars.
They found that observations of these stars are consistent with models in the hydrogen shell-burning phase, or more probably,   
in the longer lasting successive phase of core helium burning.
These stars are characterized by a deep convective envelope and a small 
convective core.
Since the density in the core is quite
 large, the buoyancy frequency can reach very large values in the 
central part, 
 with some maxima rising from the 
 steep changes in the
molecular weight. In these conditions, g modes of high frequencies can propagate and might eventually interact with p modes giving rise to modes of 
mixed character.
\begin{figure}[htb]   
\centering  
\includegraphics[draft=false,scale=0.4]{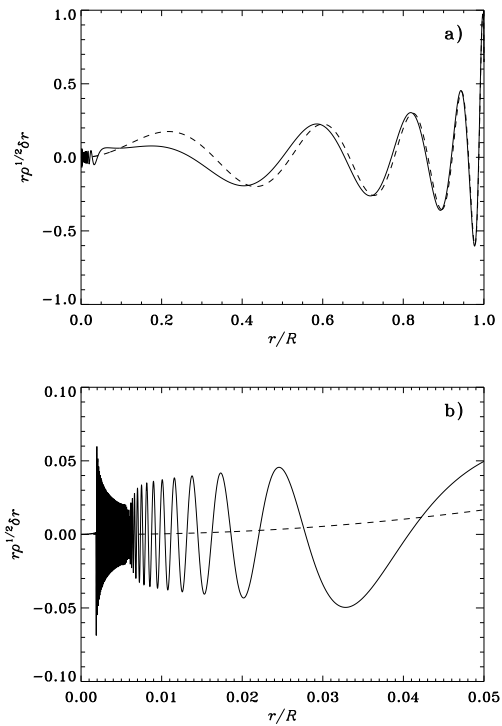}
\caption{Radial variation of eigenfunctions in a model of $\xi$ Hya. The dashed line represents a mode with l=0 and frequency $\nu=72.1\,\mu Hz$, while the solid line represents a mode with l=2 and frequency $\nu=71.3\, \mu Hz$.
Courtesy of J. Christensen-Dalsgaard.}
\protect\label{F.5}   
\end{figure}   
Fig.~\ref{F.5} shows the eigenfunctions for 
a pure radial mode and for a
mode with l=2. The eigenfunction of a mixed mode is characterized by a large p-mode like displacement amplitude in the upper layers and also g mode behaviour in the core.
Since the radiative damping scales as the third derivative of the displacement,
it appears clear that
non-radial modes are more easily damped than radial ones.
As a consequence, the spectrum of the red-giant stars is quite complex, with a 
sequence of peaks uniformly spaced, due to radial modes, which have high amplitude and hence have more probability to be detected,
and other series of peaks 
 with low amplitude, due to the mixed modes. 

\section{Asteroseismic inversion to infer the internal structure of stars}

The asteroseismic inversion is a powerful tool which allows to
estimate the physical properties of the stars,
by solving integral equations expressed in terms of the experimental data.
Inversion techniques are well known and applied with success to several 
branches of the physics, from geophysics 
to the radiation theory.
Applications to the helioseismic data have been studied extensively 
and
inversion methods and techniques have been
 reviewed and mutually compared by several authors, leading to 
 extraordinary results about the global properties of the
 Sun
(see review by, e.g., Di Mauro 2003).

The experience acquired in helioseismology on inverting mode frequency data provides a good starting point for asteroseismic inversion.
The several techniques developed for asteroseismic inversions 
support 'linear' or 'non-linear' approaches.
The linear inversion is based on the 
linearization of the equations of stellar oscillations
about a known reference model, under the assumption that perturbations are small.
This results in integral equations which
 can be used to determine the corrections which 
have to be imposed on the reference model 
in order to obtain the observed oscillation frequencies.
See the review of Basu (2003) about 
the applications and problems of linear inversions. 
The non-linear inversion based on the
 analytical resolution of the integral equations have recently been discussed and compared with the linear inversions by Roxburgh and Vorontsov (2003b).
The choice of one or the other approach is not an easy task.
The linear inversion, although is a convenient approach to study the innermost part of the stars,
 represents a questionable approximation.
The global parameters of the stars, such as mass, radius, luminosity and chemical composition are not well known, so that the structure of the model
 cannot be well constrained and hence might be very different from the observed star. In these conditions linearization  cannot be valid.
On the other hand the statistical properties, such as the propagation of errors,
 are much more defined in the linearized approach than in the non-linear inversions.

 Earlier attempts in generalizing the standard helioseismic differential methods to find the difference of structure between the observed star and a model have been applied on artificial data with encouraging results by Gough and Kosovichev (1993) and Roxburgh et al. (1998). More recently, Berthomieu et al. (2001) carried out a careful analysis of the results to be expected in inversion for stars, showing that the kernels and hence the solutions can be well concentrated only in the inner core. Patern\`o, Di Mauro and Ventura (2003) demonstrated that if a more realistic set of low degree modes is used, such as the accurate 
 observations for the Sun, the inversion allows the inference of the inner core of the stars below $0.3R_{\odot}$.
In addition, Basu, Christensen-Dalsgaard and Thompson 
(2002) demonstrated that the success of the inversion results depends strongly on the choice of variables to be inverted, and in the case of solar-type stars a preferable choice seems to be represented by the pair $u, Y$ (Basu 2003), where $u$
is the squared isothermal sound speed and $Y$ is the 
helium abundance in the convective zone. Further aspects can be also found in Thompson and Christensen-Dalsgaard (2002).

\subsection{Inversion for Procyon A}

The Procyon binary system ($\alpha$ Cmi) consists of a F5 subgiant primary and a white-dwarf secondary. Procyon A, the primary, has already attracted the attention of stellar seismologists for its proximity and brightness. 
Recently, Marti\'c et al. (2004), by observing this star with the CORALIE spectrograph,  
have been able to identify 55 p-mode frequencies in the range $250-1400\, \mu\mathrm{Hz}$, with    
harmonic degrees $l=0-2$, 
characterized by an average large frequency separation    
$\Delta \nu=53.6 \pm0.5 \, \mu\mathrm{Hz}$    
and an average small frequency separation    
$\delta \nu=5.1\pm0.5 \, \mu\mathrm{Hz}$.
These results appear to be in good agreement with those obtained 
by Eggenberger et al.~(2004a) who, from Doppler-velocity observations, 
have been able to identify 23 p~modes
in the same range of observed frequencies and with an average 
large separation of $\Delta \nu=55.5 \, \mu\mathrm{Hz}$.    
\begin{table}
 \begin{center} 
\caption{Mass, luminosity, effective temperature, radius and
logarithmic value of the iron abundance
 relative to the solar one
 for Procyon A and relative references.} 
\vspace{1em} 
\renewcommand{\arraystretch}{1.2} 
\begin{tabular}{ccc}\hline 
 Parameters &
&
References\\  \hline

$M/M_{\odot}$ & 
   $ 1.497\pm0.037$ &
  Girard et al. (2000)\\
  
$M/M_{\odot}$ & 
   $ 1.42\pm0.06$&
    Allende Prieto et al. (2002)\\
$L/L_{\odot}$&    
$7.0\pm0.2$&
Pijpers (2003)\\ 
   
$T_{\rm eff}(\mathrm{K})$&    
$6500\pm80$&
Pijpers (2003)\\   
$R/R_{\odot}$&
$2.048\pm0.025$&   
Kervella et al. (2004)\\   
   
[Fe/H]&  
$-0.015\pm0.029 $&    
  Taylor (2003)\\  
\hline    
\end{tabular}  
\label{T.3}
  \end{center} 
\end{table}   
\begin{figure}[htb]
\centering
\includegraphics[draft=false,scale=0.48]{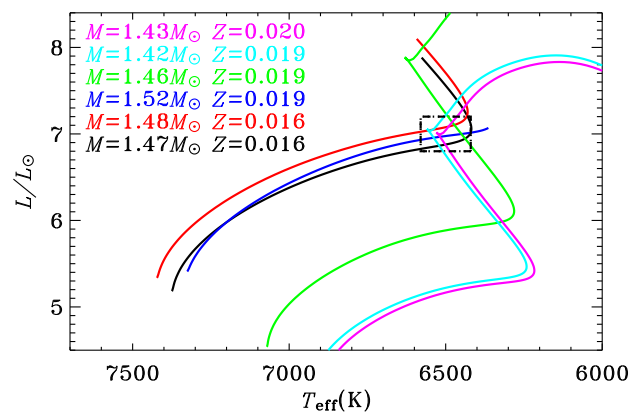}
\caption{Evolution tracks of Procyon A plotted in a HR diagram 
for several values of    
masses and metallicity chosen within the observed intervals.
The rectangle defines the one-sigma error box for   
the observed luminosity and effective temperature.}
\protect\label{F.7}
\end{figure}
\begin{table*}[t]
\caption{Basic parameters,
$ M$, $Age$, $Z$, $L$, $T_{\rm eff}$,  
the surface radius $R$, small and large separations
 for two   
models of Procyon A, computed with the OPAL EOS (1996). Model 3 and 4 represent evolutionary structure respectively  
in main-sequence and in late main-sequence phases.}\label{T.4} 
\vspace{1em} 
\renewcommand{\arraystretch}{1.2} 
\begin{center}
\begin{tabular}{ccccccccc}\hline 
 Model &  
$M/M_{\odot}$ & 
$Age$ (Gy) &
$Z$ &
 $L/L_{\odot}$ &    
 $T_{\rm eff} \; (\mathrm{K})$ &    
 $  R/R_{\odot}$ &    
 $  \delta\nu_0\, (\mu\rm{Hz})$ &
$\Delta\nu_0\, (\mu\rm{Hz})$ \\ \hline   
  3 &
$1.47$ &   
 $1.78$ &    
  $0.016$ &    
  $6.88$ &    
 $6501$  &   
 $2.07$ &
$4.2$ &
$53.6$
 \\   
   
  4 &
 $1.42$ &   
 $2.51$ &    
  $0.020$ &    
  $6.72$ &    
 $6481$  &   
 $2.05$ &
$4.2$ &
$53.6$
 \\   
   
\hline    
\end{tabular} 
\end{center}  
\end{table*}   
\begin{figure*}
\centering  
\includegraphics[draft=false,scale=0.40]{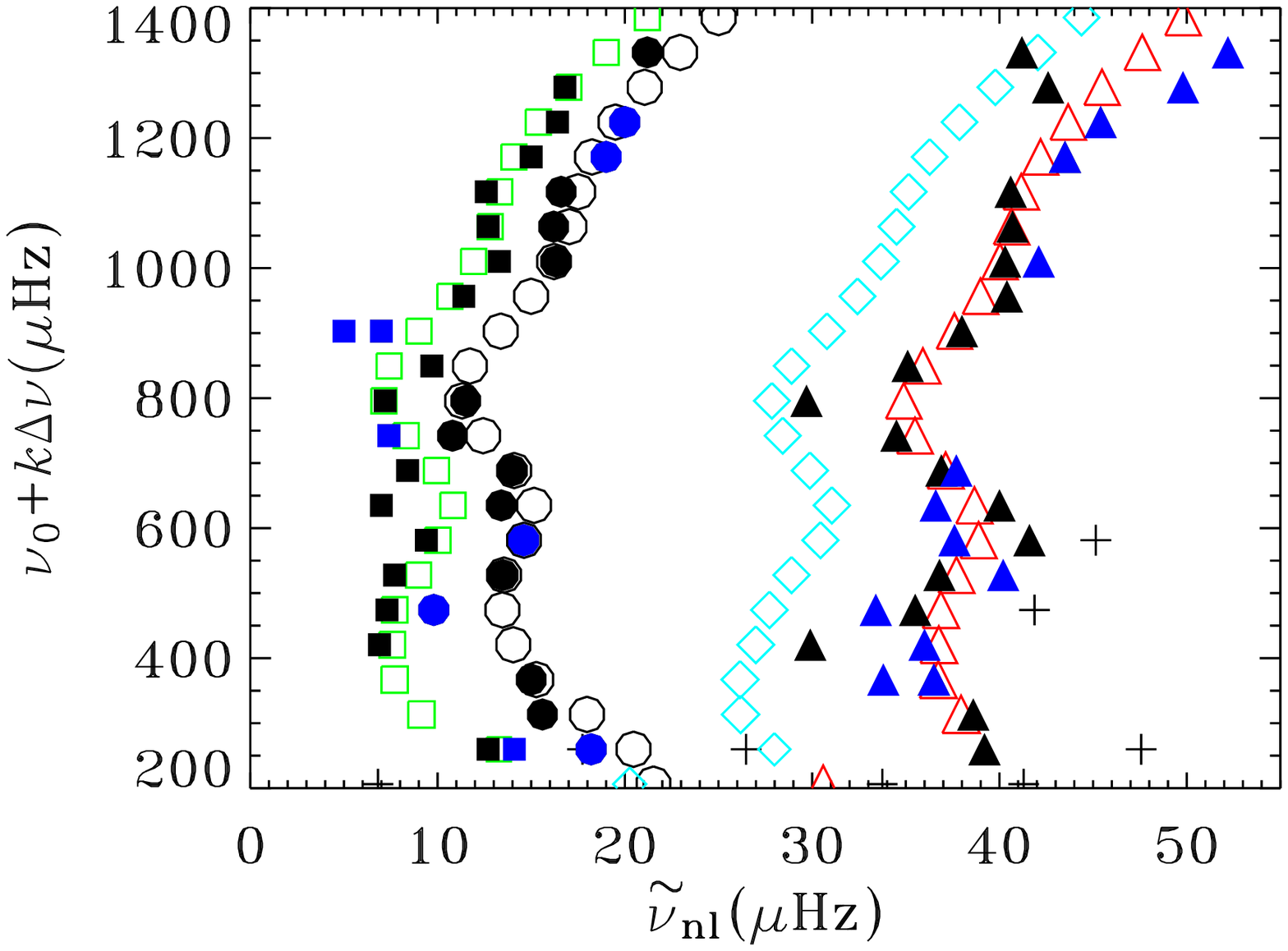}
\includegraphics[draft=false,scale=0.40]{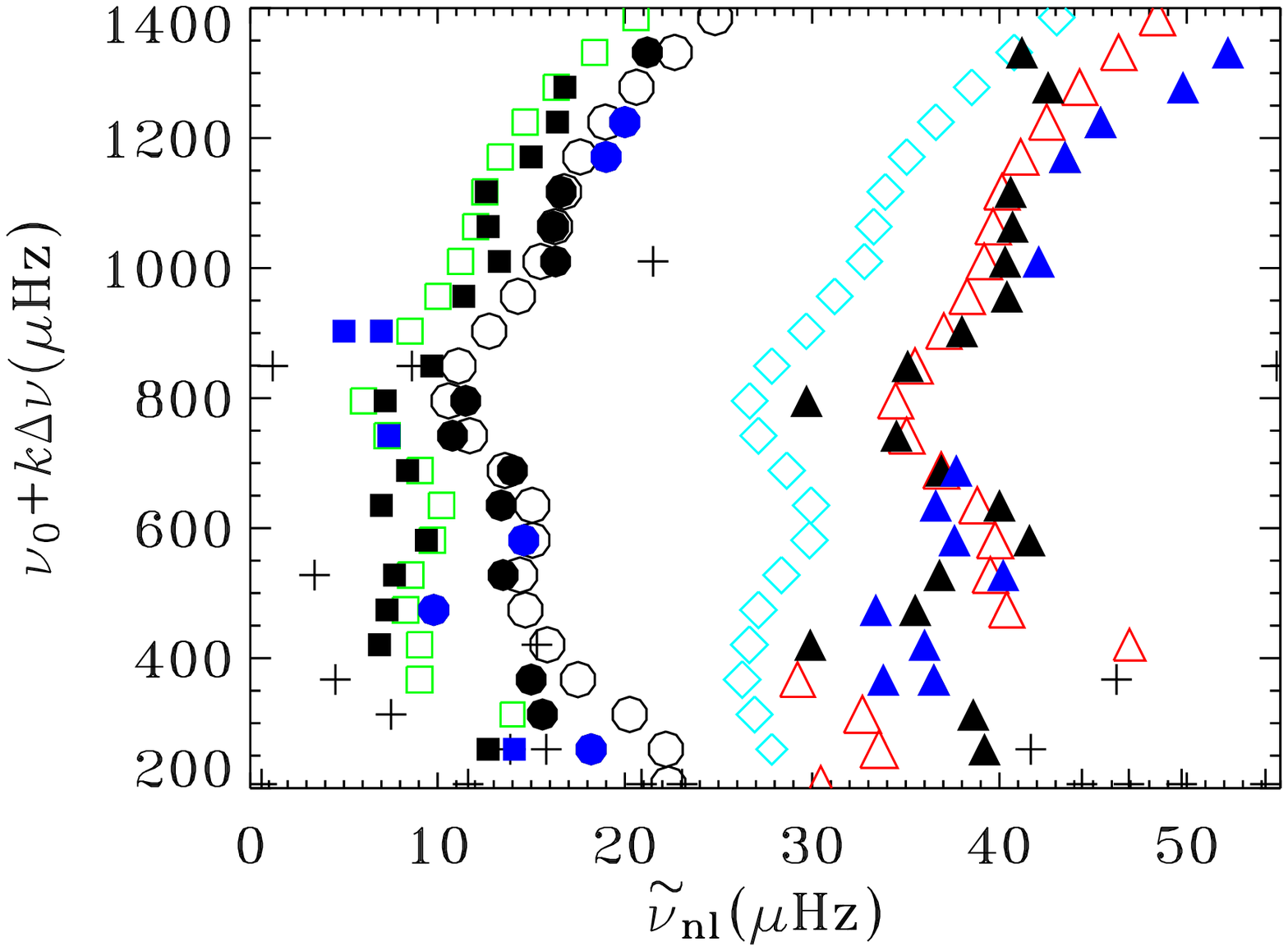}
\caption{Echelle diagrams for 
Model 3 (left panel) and Model 4 (right panel) which are respectively in the main-sequence and the late-main-sequence phases.
Open symbols represent the computed frequencies.
The filled symbols show observed frequencies by Marti\'c et al. (2004);
 blue filled symbols correspond to old identified
 frequencies given between brackets in Marti\'c et al. (2004).
Theoretical and observed frequencies are 
plotted with $\Delta \nu=53.6 \, \mu\mathrm{Hz}$ and 
a reference frequency $\nu_0=635\, \mu\mathrm{Hz}$.   
Circles are used for modes with $l=0$, triangles for $l=1$, squares
for $l=2$, and diamonds for $l=3$.   
 The size of the open symbols indicates the relative
surface amplitude of oscillation of the modes.
Crosses are employed for modes with small predicted   
amplitude (e.g. g modes).}
\protect\label{F.8}   
\end{figure*}   
The observables listed in Table~\ref{T.3} have been assumed 
to constrain a grid of stellar 
models of Procyon~A based on a recent physics, including 
the OPAL opacities (Iglesias and Rogers 1996), the \cite{Bah95} nuclear cross sections and the OPAL equation of state (Rogers, Swenson and Iglesias 1996).
Stellar models shown here have been calculated by using evolution code developed by Christensen-Dalsgaard et al. (1982), 
taking into account diffusion of helium and heavy elements from the convection zone into the adjacent radiative layers, but suppressing
 diffusion in the outer layers of the model to be sure that the helium and heavy elements were not totally drained out of the envelope.
Models shown here have been obtained for several 
heavy-element abundances,
and assuming mixing length parameter calibrated on a solar model that uses the same opacities, and an initial hydrogen mass fraction $X=0.7$

As in the case of $\eta$ Boo, it appears difficult to constrain
the evolution state of this star by considering 
only model results.
The location of the star in the HR diagram identifies Procyon A
as being in the late-main-sequence phase of core hydrogen burning if the mass 
is taken in the range $M=(1.42-1.46)M_{\odot}$.
However, 
it is also possible that Procyon~A is still in the core hydrogen-burning
main-sequence phase, if models are computed by assuming 
a mass higher by about $5\%$,
 as shown by the evolution tracks plotted
in Fig.~\ref{F.7}.\\
The problem of identifying the evolutionary state of this star can be approached
 by
studying the pulsational characteristics of the computed models.
The relevant parameters
 are given in Table~\ref{T.4} for two   
models of Procyon A selected for the pulsation analysis, 
which have surface radius, large and small theoretical separations
consistent with the observed values.
The pulsational features are clearly illustrated by looking at
the echelle diagrams for the  two different models (Fig.~\ref{F.8}).
It is evident that both the two considered models show a very good 
agreement with
the observed frequencies.
The only difference which can be noticed between the two models is
that for the model   
in the late-main-sequence phase (Model~4), the nonradial modes with $l = 1$
 show considerably   
more scatter than the radial ones.   
This aspect is associated with the avoided crossings    
which introduce a less regular structure in the frequency 
spectrum.   
On the contrary, the modes of the models in the main-sequence phase (Model~3)
show no occurrence of avoided crossing.
\begin{figure*}[htb]
\centering
\includegraphics[draft=false,scale=0.45]{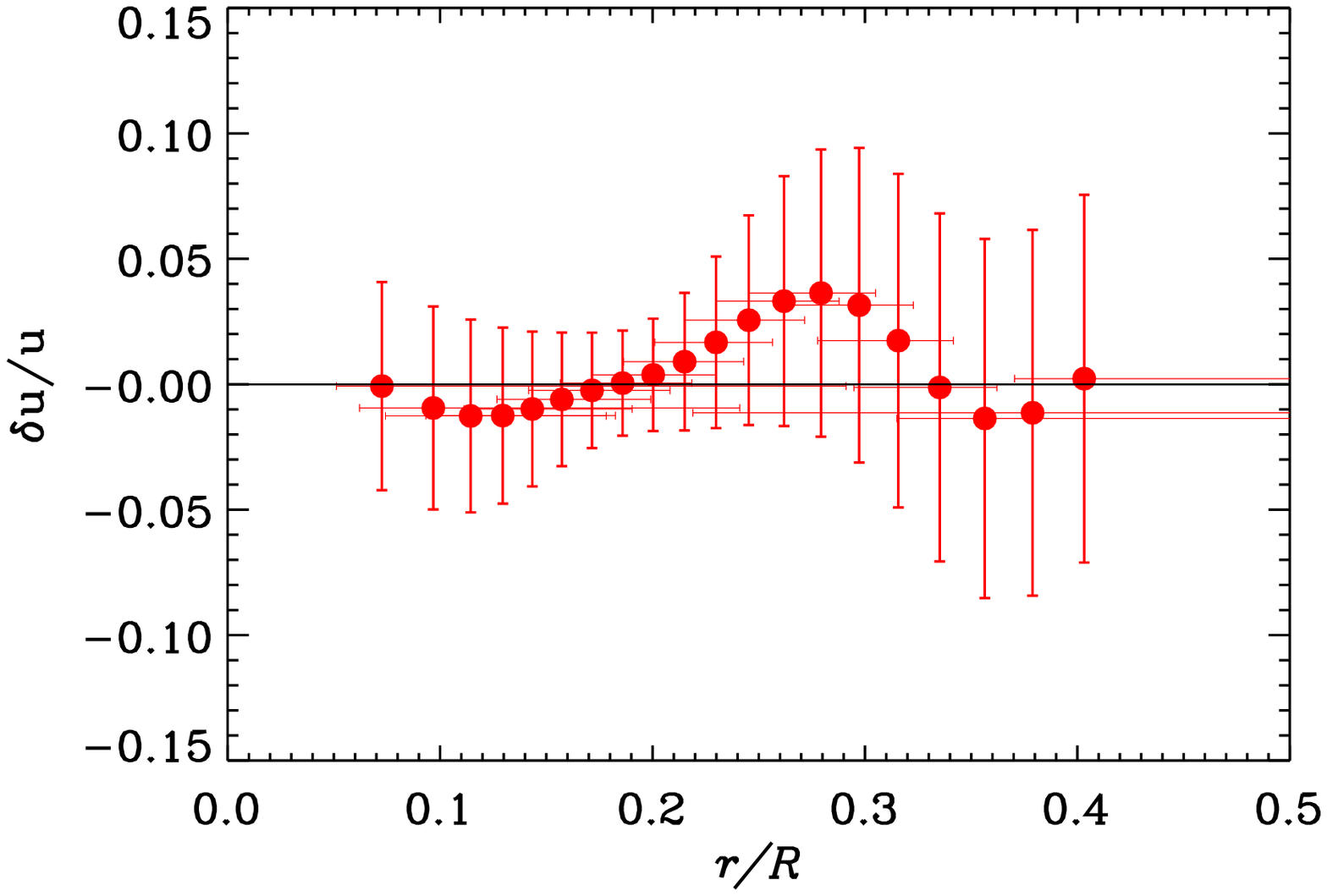}
\includegraphics[draft=false,scale=0.45]{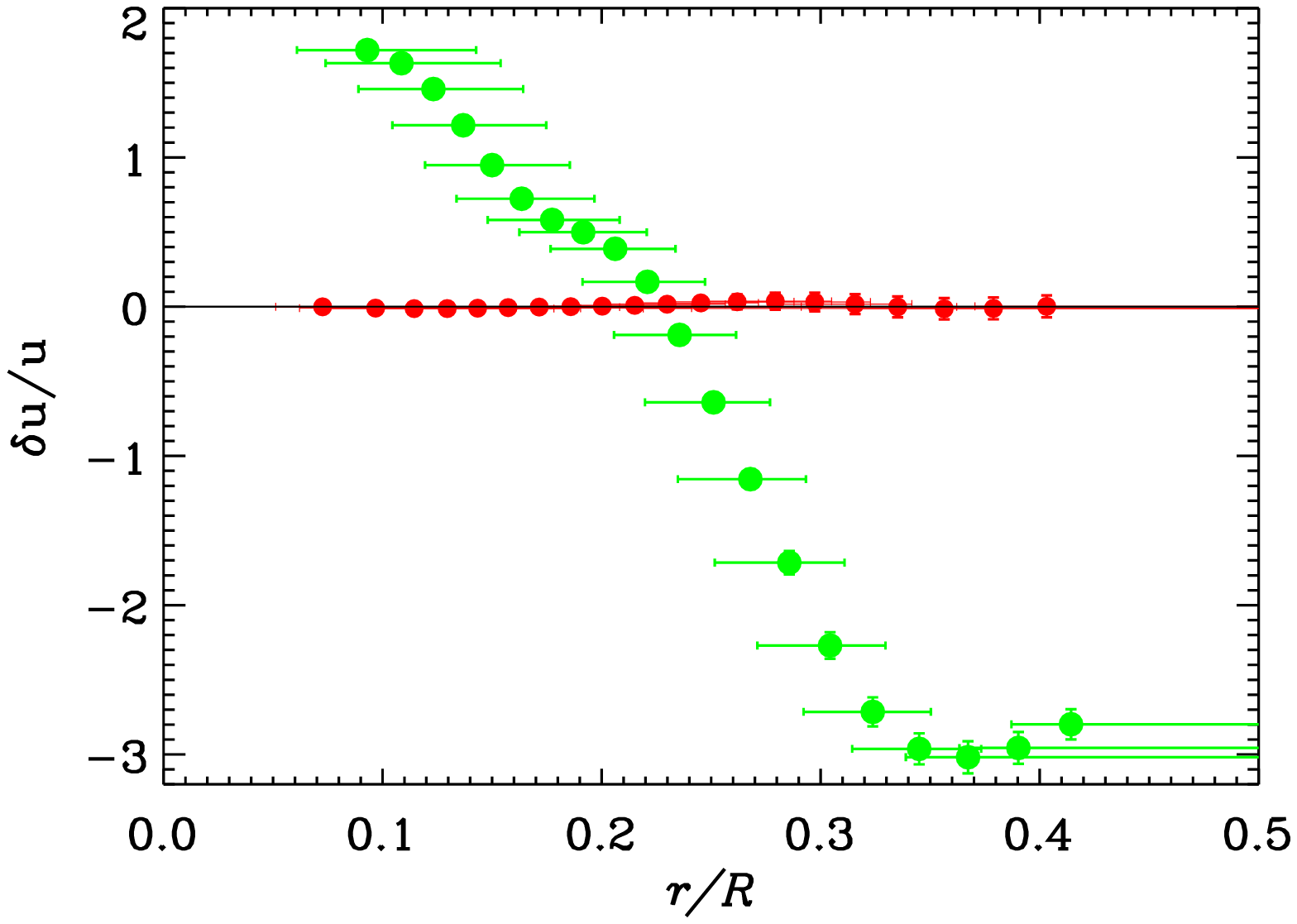}
\caption{The relative isothermal squared sound-speed differences between Procyon A and the two  
stellar models of Table~\ref{T.4} as obtained by inversion of Marti\'c et al. (2004) data. Panel on the left shows results for Model~3, panel on the right
 shows 
the comparison between the inversion results obtained for Model 3 (red dots) and those obtained for Model 4 (green dots).}
\protect\label{F.10}
\end{figure*}
Fig.~\ref{F.8} shows that the observed frequencies for $l = 1$ do not
seem to
indicate the presence of some irregularities similar to those seen for the computed
frequencies of the more evolved model.
If this were to be confirmed by more accurate sets of observed
data, Procyon A would be a star definitely identified as in the
main-sequence phase.

On considering the large amount of data available,
in order to obtain additional information on the
internal characteristics of Procyon A, it has been thought to apply
 inversion techniques to the observed frequencies of this star 
(Marti\`c et al. 2004).
  The results shown here have been obtained by
applying a linear inversion procedure. If it is assumed that the equation of state of the star is known,
the relative differences $\delta\omega_i/{\omega_i}$ between the frequencies of Procyon A and the model are related to the differences ($\delta u/u,\, \delta Y$) in squared isothermal sound speed $u=p/\rho$ and helium abundance $Y$ in the convective zone, 
between the structure of 
the star and the reference model, through the following integral equation (e.g., Dziembowski, Pamyatnykh and 
Sienkiewicz 1990):
\begin{equation}
\frac{\delta\omega_i}{\omega_i} =
 \int_0^{R}  K_{u,Y}^i \; \frac{\delta u}{u} \; {\rm d} r
+ \int_0^{R}  K_{Y,u}^i \; \delta Y \; {\rm d} r
+ \varepsilon_i  \; ,
\label{eqn:difg}
\end{equation}
where  $K_{u,Y}^i$
and  $K_{Y,u}^i$ are
the kernels calculated for each mode $i=n,l$ of the set considered.
Equation~(\ref{eqn:difg}) has been solved by applying the so called SOLA method (Pijpers and Thompson 1992, 1994) well used in helioseismology.
Fig.~\ref{F.10} shows the relative differences in $u$ between Procyon A and the two selected models 
(Models 3 and 4 of Table~\ref{T.4}) 
as functions of the 
fractional radius.
The results indicate that
the difference between Model 3 and Procyon A 
are extremely small, below the $5\%$, while Model 4 shows 
 very large deviations indicating that linerizations cannot be applied to this model, whose properties are extremely different from those of 
the observed star.

From both the inversion results and analysis of the echelle diagrams,
it can be concluded that present observations of oscillation frequencies of Procyon A seem to indicate that this star is in main-sequence phase.
It is worth to mention that Provost et al. (these proceedings)
arrived exactly to the opposite conclusion by considerations on the recent
measurements of radius by Kervella et al. (2004).
It is hoped that in the future more accurate
observations will allow a detailed test of 
stellar modelling and evolution theories of Procyon A, in order to
 identify its present evolutionary phase.

\section{Sharp Features} 

\begin{figure*}
\centering
\includegraphics[draft=false,scale=0.86]{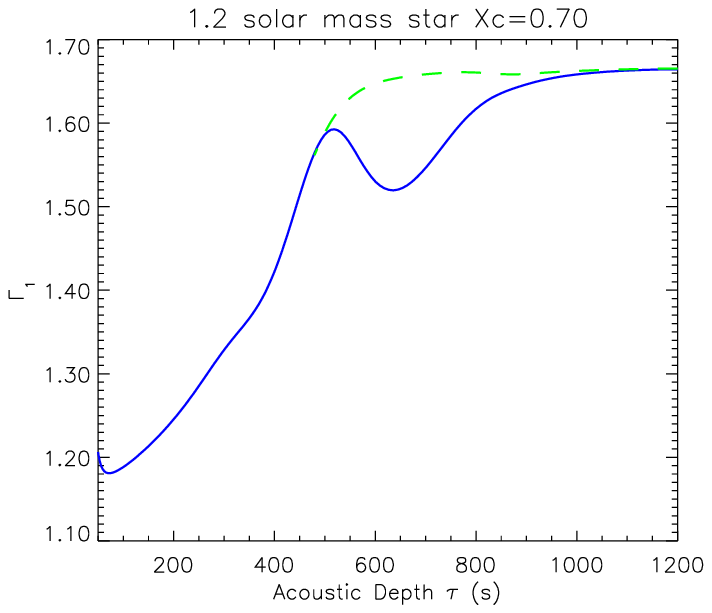}
\includegraphics[draft=false,scale=0.5]{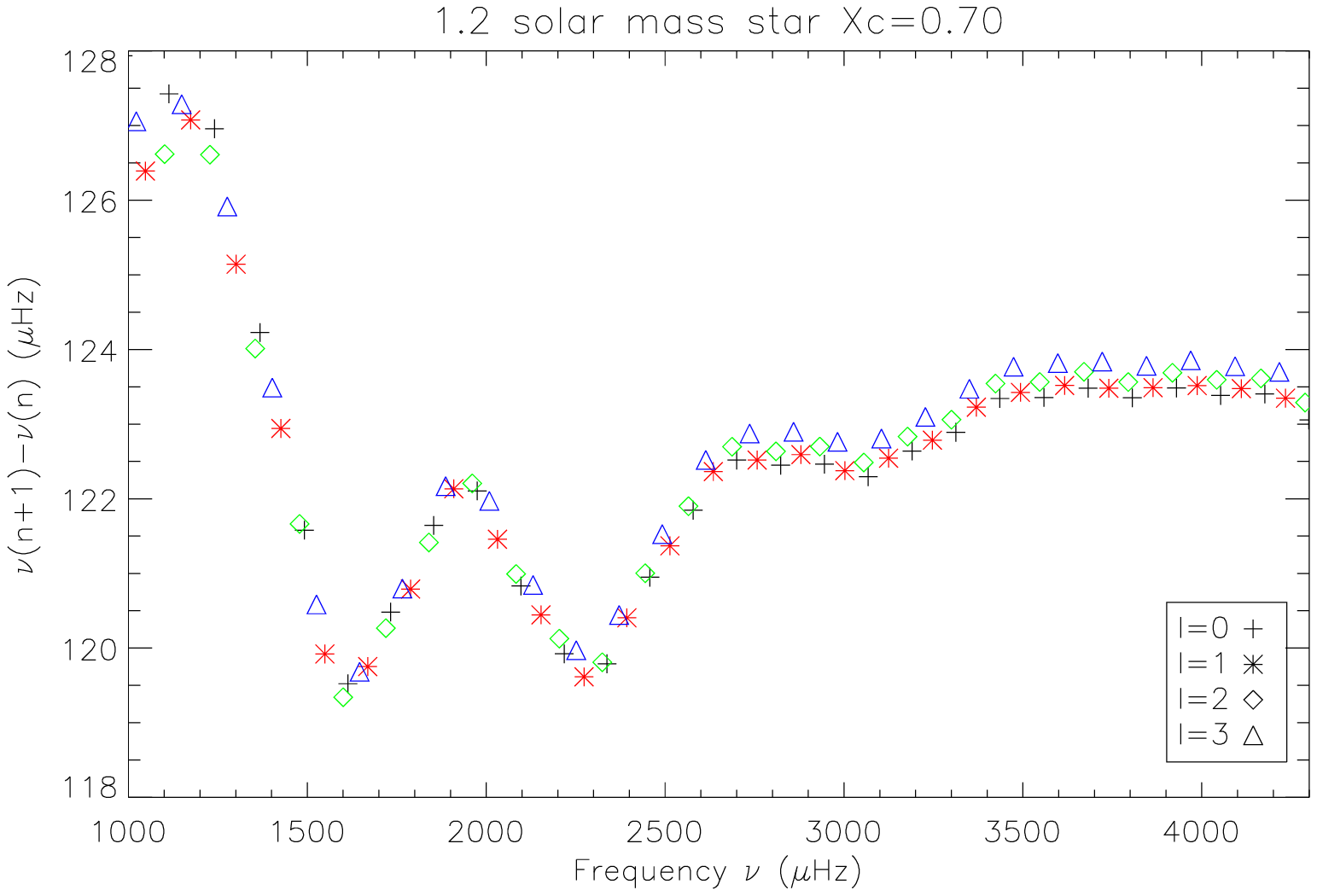}
\caption{First adiabatic exponent $\Gamma_1$  for a ZAMS stellar model of $1.2M_{\odot}$ (left panel), where the dip corresponds to the second helium ionization region. The relative oscillatory behaviour it is shown
on the right panel where the large frequency separations is plotted as function of the frequency.}
\protect\label{F.11}
\end{figure*}

Another important property of the oscillation spectra
is that sharp variations localized at certain acoustic depth in the structure of pulsating stars produce a distinctive quasi-periodic signal in the  
frequencies of oscillation.
The characteristics of such signal are related to the location and thermodynamic properties of the layer where the sharp variation occurs.
Sources of sharp variations are the borders of convection zones and regions of rapid variation in the first adiabatic exponent $\Gamma_1$, such as the one that occurs in the region of second ionization of helium.
Fig.~\ref{F.11} shows the variation of $\Gamma_1$ in the upper layers for a ZAMS model of a star of $1.2M_{\odot}$ and its
relative periodic signal as seen in the 
large separation.

A general expression (valid for low degree modes) for the signal generated is
 of the form (e.g. Monteiro, Christensen-Dalsgaard and Thompson 1994, 1998, 2000; Roxburgh and Vorontsov 1994):
\begin{equation}
\delta\omega_{n,l}\simeq A(\omega)\cos(2\omega_{n,l}\tau_d+\phi_0)\, ,
\label{sharp}
\end{equation}
where $A(\omega)$ is an amplitude as function of frequency $\omega$,
 which depends on the properties of the sharp variation; $\phi_0$ is a constant phase; $\tau_d$ is the acoustic depth of the feature.
Several attempts have been tried in order to isolate the generated oscillatory components directly from the frequencies of oscillations or from linear combination of them (large separations, second differences, etc). 
The common approach consists in removing a smooth component from 
 the frequencies and 
to fit the residual signal to a theoretical expression, like that
of Eq.~(\ref{sharp}),
which is related to the properties of the sharp feature.

This method can be applied, for example, to determine the properties of the base of the convective envelope (Monteiro, Christensen-Dalsgaard and Thompson 
2000; Ballot, Turck-Chi\`eze and Garc\'{\i}a 2004) and in particular to put limits on the extension of the 
convective overshoot (Monteiro, Christensen-Dalsgaard and Thompson 2002),
or to investigate the border of the convective core 
(Mazumdar and Antia 2001; Nghiem et al. 2004). 
But in particular, this peculiar property of the oscillation frequencies
can be used to infer the helium abundance in the
 stellar envelope, by 
 studying the variation of $\Gamma_1$ in the region of 
second ionization of helium.
 Several authors have tried to develop or refine their own fitting function 
to infer the helium abundance (Lopes et al. 1997; Monteiro and Thompson 1998; P\'erez Hern\'andez and Christensen-Dalsgaard 1998; Miglio et al. 2003; Basu et al. 2004 and these proceedings; Houdek and Gough in these proceedings).
Unfortunately this technique cannot be
applied to stars with mass larger than $1.4M_{\odot}$, owing to the 
contamination in the oscillatory signal coming from the first helium ionization zone.
In fact, one of the main problem in the application of such an approach rises from the fact that signals
coming from different sharp features in the interior of the star might 
overlap generating a complex behaviour, as it was noticed by Mazumdar and Antia (2001). A similar effect was also considered by Montgomery, Metcalfe and Winget (2003) in the case of g modes observed in white dwarfs.
 Very recently Miglio and Antonello (2004) have shown that in the solar-type stars the problem of distinguishing between a signal generated near the core or near the surface in the star can be easily solved by considering modes of different degree or different seismic indicators. Unfortunately this method cannot successfully be applied in the case of the white dwarfs.
At the moment, L. Mantegazza et al. (private communication) is working on a new technique based on the Principal Component Analysis (Golyandina et al.
2001; Ghil et al. 2002)  in order to isolate all the different oscillatory components directly from the oscillation frequencies.

It was already pointed out in Miglio et al. (2003) that a more general analysis of the effect of the bump in $\Gamma_1$ can be obtained by performing an inversion of the observed frequencies, by using the integral equation by Gough and Thompson (1991).
At the moment the method has been tested only in order to find differences
between a reference equilibrium model and a fictitious model which differs from the reference one only because
$\Gamma_1$ has been smoothed in the region of second helium ionization.
In this case, the inversion of the
 artificial data set of modes with $0\leq l\leq 3$ is able to reconstruct the variations in $\Gamma_1$ for models with masses from low up to $2M_{\odot}$,
if the inverted set includes also few modes 
with low radial order ($n=4-8$), whose kernels 
seem to have considerable weight in the upper regions.

\section{Theoretical results on other pulsating stars}
\label{S.WD}

\subsection{White dwarfs}

The results obtained by applying seismic techniques to white dwarfs represent probably the biggest success of asteroseismology (e.g. Vauclair 1997; Kawaler 1998; O'Brien 2003). In fact, they represent the stars, other than the Sun,
 in which the largest number of
 oscillation frequencies has been detected. The white-dwarf stars (WDs) represent the most common end point of stellar evolution.
Their physical structure is very simple: a degenerate carbon-oxygen core and an extremely 
thin, mostly non-degenerate, surface layer of pure H or He. 
Pulsations in WDs represent a short-lived phase during their evolution: the majority of them do not show pulsations except in very narrow regions of instability located along their cooling 
track in HR-diagram: the planetary nebulae hot nuclei (PNN), the hot DO (H deficient pre-white-dwarfs), the warm DB 
(WDs with He rich photospheres) and the cool DA (WDs with H rich 
photospheres) stages.
WD variables are mainly multi-periodic pulsators in which high-order
g-modes of low angular degree are excited, but the mechanism triggering the pulsations is still matter of discussion (e.g., Cox 2002).

Asteroseismology has been applied to identify the structural differences between different types of white dwarfs, and there is hope that it will
permit in the future also the identification
of the various origins.
In fact, while theoretical models can be easily calculated without major efforts, it is still not possible to say which stellar structure might represent the precursor of each of the several types of white dwarfs.

Asymptotic theory, which gives a clear prediction of period spacing for these stars, has helped to determine the total mass, the rotation rate, the magnetic field strength and even the mass of the outer layers of some selected targets.
In particular, since the rate of evolution of white dwarfs depends primarily on the thickness of their surface layers, it might be possible to measure
the galactic age through the luminosity function (e.g. Wood and Oswalt, 1998).
However, the most interesting characteristics studied by asteroseismology are the structure and composition of the C/O cores of white dwarfs and the evidence for
crystallization of the cores in the most massive stars (DAV). 
As shown by Montgomery and Winget (1999), crystallization leaves a typical signature on the period spectrum with an amplitude which increases as the fraction of crystallized matter increases.
Recently, two recent debates have been opened on the possibility to measure 
the fraction of
crystallized matter (see Metcalfe, Montgomery and Kanaan 2004; Brassard and Fontaine these proceedings) and the $^{12}C(\alpha, \gamma)^{16}O$ reaction rates in white dwarfs (see
Metcalfe, Salaris and Winget 2002; Fontaine and Brassard, 2002) by studying 
their pulsational properties .

Some WD variables have, or are expected to have, spectra rich enough in modes to provide 
useful constraints on their rotational rates. 
Attempts to derive the internal rotational profiles by using the observed splitting and adopting forward calculations and inversion techniques have been 
presented in several works (e.g. Kawaler, Sekii and Gough 1999, Vauclair et al. 2002).  

Finally, another asteroseismic achievement is represented by the prediction
 that, during the pre-white-dwarf stage, the dominant energetic process is the neutrino cooling. Since evolutionary changes affect the
pulsation periods, the detection of secular period changes in the hottest
WD pulsators allow to measure their cooling rates, 
providing experimental constraints on the theories of neutrino emission in dense plasmas and thermal neutrino production rates (O'Brien and Kawaler 2000).

\subsection{$\delta$ Scuti stars}

$\delta$-Scuti stars are typically population I, A and F main-sequence or slightly post 
main-sequence objects of about $2-2.5\,M_\odot$. 
They are
 located in the lower part of the classical Cepheid instability strip  where the $\kappa$-mechanism is expected to drive low-order radial and non-radial p-modes, modes with a mixed p- and g-mode character, and possibly g-modes. 
Most of the $\delta$-Scuti stars are moderate or rapid rotators with surface velocities up 
to $100-200\,{\rm km\,s^{-1}}$.

The identification of the oscillation modes is a very complex task for these stars, since the asymptotic theory does not apply to the excited modes (low-order p-modes).
However, the main problem depends on the scarcity of observed modes,
among the hundreds of excited pulsations predicted by theory.
 Rapid rotation and possible differential rotation produce a rather complex power spectrum, making more difficult the identification of modes without additional information.\\
Thus, at the present, asteroseismology is able to put only some constraints on the internal structure of $\delta$ Scuti stars.
Many interesting and comprehensive reviews on asteroseismology of $\delta$-Scuti stars can be found in the recent literature (e.g., 
Breger and Montgomery 2000; Goupil and Talon 2002; Paparo, these proceedings).

\subsection{Rapidly oscillating Ap (roAp) stars}

The rapidly oscillating Ap (roAp) variables, a subgroup of the chemically peculiar A-type stars, are H core-burning stars of mass $\simeq$ 2 M$_{\odot}$, characterized  by strong dipole magnetic fields 
of the order of a few kG, which are the most intensive fields in main-sequence stars. 
Their position in the H-R diagram overlaps the $\delta$-Scuti star instability
strip (see Fig.~1), but the two groups of variables differ significantly in the behaviour of the excited modes. The roAp stars pulsate in high-order ($n\ge 20$), low-degree p-modes. Most of the roAp stars pulsate in almost pure dipole ($l=1$) modes.
The reason why stars with similar luminosity and effective temperature may show so different pulsation characteristics is not completely understood, even though it is clear that the strong magnetic fields could play an important role in selecting the excited modes.    

The parallax measurements have provided a primary opportunity
to check independently the predictions of roAp stars seismology.
The results pointed out the tendency for seismic predicted values to be systematically slightly smaller than the  Hypparcos ones (Matthews et al. 1999).

The observed pulsation properties of roAp stars were originally explained in terms 
of the so-called oblique pulsator model (Kurtz 1982), in which pulsation and magnetic axes are mutually aligned, but tilted with respect to the rotation axis.  The oblique pulsator model has been improved in order to better reproduce the observed power spectra of roAp stars (e.g. Dziembowski and Goode 1996) and more recently revisited
 by Bigot and Dziembowski (2002) which found out that
the
 rotational effects seem to be so strong
to result in a pulsation axis inclined with respect to both the rotation and the magnetic axes. 
However, the new theory still needs to be tested.

The excitation mechanism for the roAp stars is still an unresolved problem, although extensively debated over the years. The $\kappa$-mechanism in the He II ionization region which drives $\delta$-Scuti star oscillations has been demonstrated to be inadequate for exciting high-frequency modes in roAp stars, even
though the two groups of variables share the same location in the H-R diagram. A great deal of alternative models were then proposed, each of them studying various effects which could affect the $\kappa$-mechanism giving rise to mode instability, such as a reduced convective efficiency due to the action of  the magnetic field (Balmforth et al 2001), settling of He (Dolez and Gough 1982), presence of a stellar wind (Dolez, Gough and Vauclair 1988) or a chromosphere (Gautschy, Saio, Harzenmoser 1998).

\section{Conclusion}

Asteroseismology constitutes a unique approach to the direct investigation of stellar structure and evolution which can significantly improve our knowledge of astrophysics, giving
information on the structure, dynamics and evolutionary stage of the stars.
Asteroseismology is still far from the great results of helioseismology, but
it is clear that
its success will depend on the amount and quality of data collected by the ground-based observatories, the significant improvement in the accuracy expected from the future space missions, and the progresses in the development of the application to the stars 
of techniques for the data inversion and study of the 
effects of sharp features on the oscillation frequencies.  

\section*{Acknowledgments} 
I am grateful to the SOC and in particular
to Sarbani Basu, for the kind invitation
and to ESA for the financial support
which allowed me to participate to the {\it SOHO14/GONG2004} conference
with the present contribution.
I am very grateful to J. Christensen-Dalsgaard, M. Cunha, 
J. Daszy\'nska-Daszkiewicz, A. Miglio, J. Montalban,
M. J. P. F. G. Monteiro and L. Patern\`o
 for fruitful discussions, comments and help to realize
and improve this review.
Finally, I wish to thank D. Cardini for the kind hospitality 
at the institute IASF of Rome where I had the possibility to accomplish 
the preparation of the present paper.

\end{document}